\documentclass[conference,10pt]{IEEEtran}
\IEEEoverridecommandlockouts
\usepackage[T1]{fontenc}
\usepackage[utf8]{inputenc}
\usepackage{cite}
\usepackage{comment}
\usepackage{graphicx}
\usepackage{todonotes}
\usepackage{caption}
\usepackage{subcaption}
\usepackage[hidelinks]{hyperref}
\usepackage{orcidlink}
\usepackage{paralist}

\usepackage{lipsum}

\begin{document}
\title{The CAPSARII Approach to Cyber-Secure Wearable, Ultra-Low-Power Networked Sensors for Soldier Health Monitoring
\thanks{This work has been funded by the European Union via the European Defence Fund project CAPSARII under grant number 101168102.
Views and opinions expressed are however those of the author(s) only and do not necessarily reflect those of the European Union or the European Defence Fund.
Neither the European Union nor the European Defence Fund can be held responsible for them.}
}
%
%
\author{\IEEEauthorblockN{
Luciano Bozzi\IEEEauthorrefmark{1}\orcidlink{0009-0009-3920-3546},
Christian Celidonio\IEEEauthorrefmark{1},
Umberto Nuzzi\IEEEauthorrefmark{1}\orcidlink{0009-0002-0329-7287},
Massimo Biagini\IEEEauthorrefmark{1}\orcidlink{0009-0004-5655-1917},
Stefano Cherubin\IEEEauthorrefmark{2}\orcidlink{0000-0002-5579-5942},\\
Asbjørn Djupdal\IEEEauthorrefmark{2}\orcidlink{0009-0002-3571-2828},
Tor Andre Haugdahl\IEEEauthorrefmark{2}\orcidlink{0009-0002-9876-5214},
Andrea Aliverti\IEEEauthorrefmark{3}\orcidlink{0000-0002-2950-0231},
Alessandra Angelucci\IEEEauthorrefmark{3}\orcidlink{0000-0002-1957-2231},
Giovanni Agosta\IEEEauthorrefmark{3}\orcidlink{0000-0002-0255-4475},\\
Gerardo Pelosi\IEEEauthorrefmark{3}\orcidlink{0000-0002-3812-5429},
Paolo Belluco\IEEEauthorrefmark{4}\orcidlink{0000-0001-6877-9370},
Samuele Polistina\IEEEauthorrefmark{4}\orcidlink{0009-0002-5179-7040},
Riccardo Volpi\IEEEauthorrefmark{5}\orcidlink{0000-0003-4485-9573},
Luigi Malag{\`o}\IEEEauthorrefmark{5}\orcidlink{0000-0002-2392-5108},\\
Michael Schneider\IEEEauthorrefmark{6},
Florian Wieczorek\IEEEauthorrefmark{6}\orcidlink{0000-0002-9825-2569},
Xabier Eguiluz\IEEEauthorrefmark{7}\orcidlink{0000-0002-3314-2465}
}
\IEEEauthorblockA{\IEEEauthorrefmark{1} Sea Sky Technologies, Rome, Italy:
\{l.bozzi, c.celidonio, u.nuzzi, capsarii\_ext\_1\}@seaskytech.it
}
\IEEEauthorblockA{\IEEEauthorrefmark{2} NTNU, Trondheim, Norway:
\{stefano.cherubin, djupdal, tor.a.haugdahl\}@ntnu.no
}
\IEEEauthorblockA{\IEEEauthorrefmark{3} Politecnico di Milano, Milano, Italy:
\{andrea.aliverti, alessandra.angelucci, giovanni.agosta, gerardo.pelosi\}@polimi.it
}
\IEEEauthorblockA{\IEEEauthorrefmark{4} LWT3, Milano, Italy:
\{paolo.belluco, sam.polistina\}@lwt3.com
}
\IEEEauthorblockA{\IEEEauthorrefmark{5} QUAESTA AI, Cluj-Napoca, Romania:
\{riccardo.volpi, luigi.malago\}@quaesta.ai
}
\IEEEauthorblockA{\IEEEauthorrefmark{6} BORN GmbH, Dingelst\"{a}dt, Germany:
michael.schneider@born-germany.de, florian.wieczorek@smarttexhub.com
}
\IEEEauthorblockA{\IEEEauthorrefmark{7} IKERLAN, BRTA, Arrasate-Mondragón, Spain:
xeguiluz@ikerlan.es
}
}

\maketitle              
\begin{abstract}
The European Defence Agency's revised Capability Development Plan (CDP) identifies as a priority improving ground combat capabilities by enhancing soldiers' equipment for better protection. The CAPSARII project proposes an innovative wearable system and Internet of Battlefield Things (IoBT) framework to monitor soldiers' physiological and psychological status, aiding tactical decisions and medical support.
The CAPSARII system will enhance situational awareness and operational effectiveness by monitoring physiological, movement and environmental parameters, providing real-time tactical decision support through AI models deployed on edge nodes and enable data analysis and comparative studies via cloud-based analytics.
CAPSARII also aims at improving usability through smart textile integration, longer battery life,  reducing energy consumption through software and hardware optimizations, and address security concerns with efficient encryption and strong authentication methods. This innovative approach aims to transform military operations by providing a robust, data-driven decision support tool.

\end{abstract}
\begin{IEEEkeywords}
EDF (European Defence Fund), IoBT, Edge AI, Wearable systems.
\end{IEEEkeywords}


\section{Introduction}

The European Defence Agency has recently revised its Capability Development Plan (CDP)~\cite{edf2023cpd} to reach new levels of preparedness and cooperation among the member countries.
Among the 11 priorities identified in the CPD, improving ground combat capabilities requires enhancing the equipment of individual soldiers to obtain better protection.
Fine-grain awareness of the physiological and psychological status of each soldier can be a key tool for making sound tactical decisions and providing life-saving information to medical support services.

To this end, CAPSARII proposes the development of an innovative wearable system and related ICT framework, designed specifically for military staff, which aligns with the concept of the Internet of Battlefield Things (IoBT)~\cite{russell2018internet}.
IoBT refers to a network of interconnected devices with sensing capabilities on the battlefield and that seamlessly share data and insights to enhance situational awareness and operational effectiveness.
This wearable system could represent a key component to transform the monitoring of soldiers' psychophysical conditions and the evaluation of their performance, both in combat and in training scenarios.

While wearable devices are currently employed in civilian contexts such as healthcare, sports, and industry, defence applications require an enhanced range of capabilities, as well as a wholly different set of non-functional requirements~\cite{ometov2021survey}.
The great range of sensors required by defence applications impacts both the computing and the radio transmission, increasing the energy required for the system to work, and therefore reducing the battery lifetime, which is a critical aspect for extended functionality:
while civilian devices with limited purposes can be tailored to survive for days on battery, and medical devices can be built sacrificing wearability for extended functionality, such compromises are unacceptable in the battlefield.

In this paper, we outline the key objectives, components, and technologies that will make this IoBT-enabled wearable system a crucial and transformative asset for modern military forces.
By employing advanced prescriptive and predictive tools based on AI, CAPSARII empowers the military chain of command with a data-driven decision support tool, which is designed to continuously improve with usage and be robust in different battle conditions.

CAPSARII aligns its scientific and technical activities along five strategic objectives:
\begin{inparaenum}
\item enabling the monitoring of physiological, movement and environmental parameters;
\item enabling tactical decision support at the edge;
\item enabling data update, analysis and comparative studies;
\item preserving the security and confidentiality of the information;
\item ensuring the usability of the wearable systems by maximising its battery lifetime and improving the integration within smart textiles.
\end{inparaenum}

To carry out this ambitious research proposal, CAPSARII is powered by a consortium including top European academic institutions such as Politecnico di Milano (POLIMI) and the Norwegian University of Science and Technology (NTNU), a leading technology transfer centre, Ikerlan (IKER), and a set of SMEs covering the supply chain of secure IoBT (SeaSkyTechnologies), data analytics for bio-signals (LWT3), smart textiles (BORN), and AI and Machine Learning (QUAESTA).
SeaSkyTechnologies (SST), a consulting company and system integrator with significant expertise in the defence sector, leads the consortium.

\section{Use Case Scenarios in Air and Land Domains}

The CAPSARII concept and mechanisms will be evaluated in two use case scenarios representative of the general problem: \emph{pilot fatigue} and \emph{warfighters in harsh environments}.
The two use cases address the air and land domains, and are also suitable for approaching and evaluating the use of federated learning through the implementation of a shared model among the users in which the training phase could be performed using decentralised data involving EU participants of different member states, e.g., during joint exercises among EU countries and NATO alliance that regularly take place to highlight Operational and Tactical capabilities of Special Operations Forces.

\begin{figure}
	\begin{subfigure}{0.49\textwidth}
	\includegraphics[width=\linewidth]{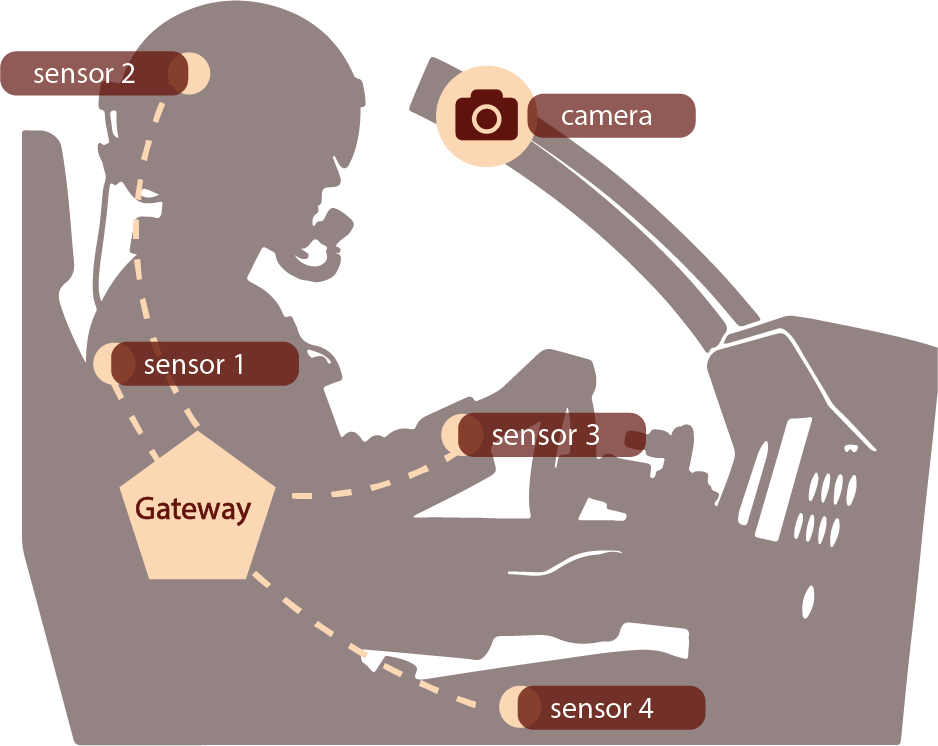}
	\caption{\label{fig:uc1}Use Case 1: Pilot Fatigue}
	\end{subfigure}
	\quad
	\begin{subfigure}{0.5\textwidth}
	\includegraphics[width=\linewidth]{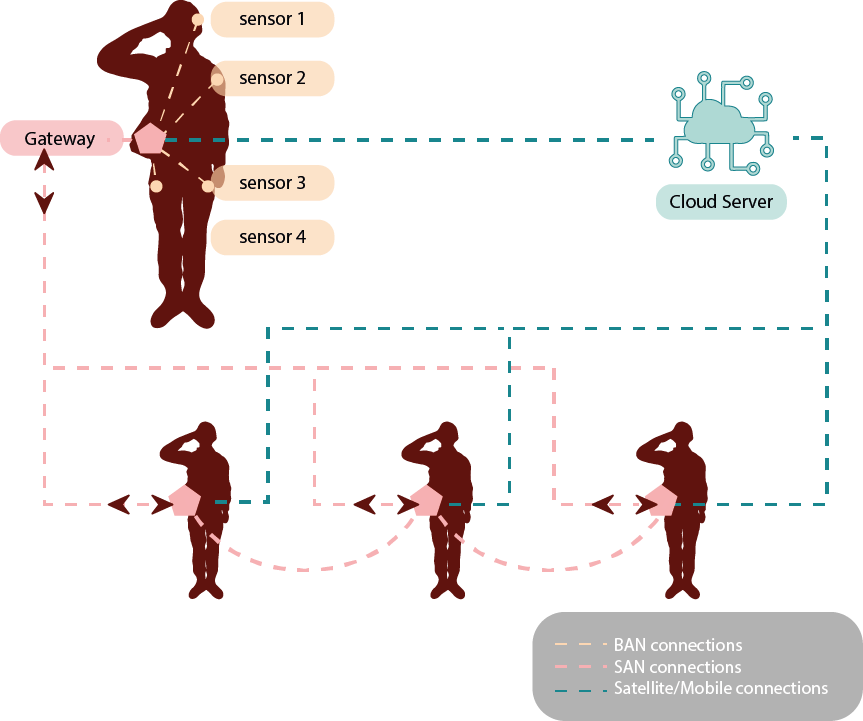}
	\caption{\label{fig:uc2}Use Case 2: Warfighters}
	\end{subfigure}
	\caption{\label{fig:uc}CAPSARII Use Cases}
\end{figure}

\paragraph{Pilot Fatigue}
Pilot fatigue monitoring, an implementation of which is shown in Figure~\ref{fig:uc1}, is receiving increasing attention, as sleep deprivation can lead to severe impacts on aviation safety as well as service member health and readiness~\cite{gregory2010pilot,samel2004sleep}.
Alertness and the ability to perform are related to two basic neurophysiological forces: the body’s circadian pacemaker (or biologic clock) and the drive or need for sleep (based on the length of previous wakefulness).
Some countermeasures have been identified, e.g., exposure to cold air, avoidance of night/morning work, naps, caffeine, melatonin~\cite{caldwell2001impact}.
Personalised biomathematical models of fatigue, incorporating an individual's traits, can provide guidance to crew members on their expected level of fatigue at a given time, which could be used to improve sleep management strategies and to apply personalised fatigue countermeasures~\cite{crs2022,authority2014biomathematical}.

\paragraph{Warfighters in Harsh Environments}
Real-Time Physiological Status Monitors (RT-PSM)~\cite{friedl2016real} are an emerging class of modern military wearable technologies, providing individual soldiers and their leadership with actionable physiological status information needed to ensure individual and squad health and performance/readiness.
Such systems usually employ a Body Area Network (BAN)~\cite{telfer2020open} to convey the information, but an emerging concept of Squad Area Network (SAN)~\cite{lai2009digital} enables intra-squad communications.
This concept, shown in Figure~\ref{fig:uc2} allows monitoring, and possibly predicting, the physiological status and the health conditions of single individuals and of whole teams within the SAN.
By coupling the physiological status and the individual's movement with environmental conditions, it will be possible to learn the causes and possibly to prevent accidents due to stress or fatigue.

These use cases and their implementation pose several challenges in terms of Low-Power Embedded AI, real-time processing, low energy dissipation, federated learning, communication reliability, performances and security. 
The CAPSARII methodology could represent a viable approach to address the topic of monitoring stress, fatigue, and physiological factors on the field alongside physical health and environmental situation at different layers of the picture representation and decision-making process.

\section{Technical Approach}

\begin{figure*}
\begin{center}
	\includegraphics[width=\textwidth]{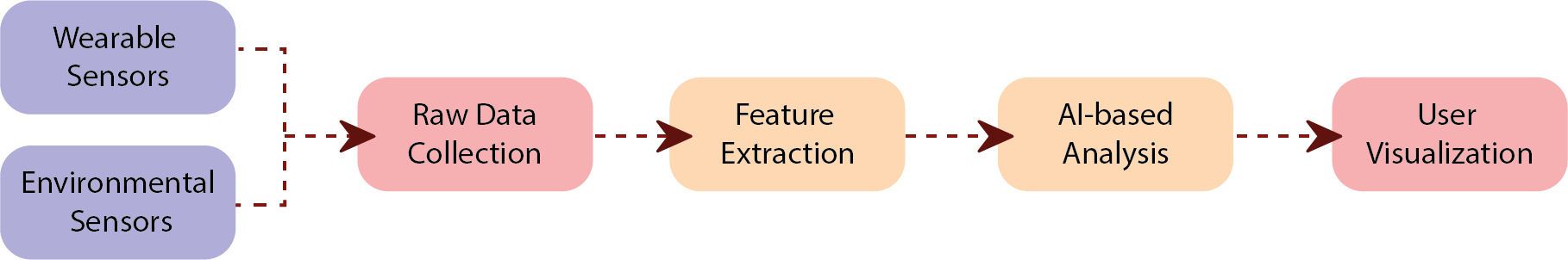}
	\caption{\label{fig:pipe}The CAPSARII data collection and analysis architecture}
\end{center}
\end{figure*}

The CAPSARII project is founded on the concept of Internet of Battlefield Things (IoBT), which interconnects military-grade equipment into a protected IoT computing architecture.
This system, depicted in Figure~\ref{fig:pipe} is envisioned as a continuous, interconnected process, ensuring efficient data flow and analysis to and from the wearable devices.

\paragraph{Data collection to the edge}
The wearable devices in the CAPSARII system collect real-time data on various physiological, movement and environmental parameters, and operational metrics using a mesh network.
This data is then processed on edge-computing systems, ensuring rapid and localised analysis.
Edge analysis involves the initial processing and evaluation of data near the source, on wearable devices or within the near network.
This localised analysis enables rapid decision making and minimises latency in the acquisition of critical information.

\paragraph{From the edge to the server}
After initial analysis at the edge, relevant data is transmitted to a centralised server for further processing and storage.

\paragraph{Cloud-based analysis}
Once at the external server, the CAPSARII IoBT system leverages cloud-based analytics to perform in-depth evaluation.
The cloud can extend its analytical capabilities through a collaborative approach, interweaving different cloud resources from different EU countries.
This collaborative cloud-based analytical framework will be able to enable comprehensive assessments, data correlation, and meaningful pattern extraction on a larger scale, facilitating advanced analysis and pattern formation.

\paragraph{Visualisation}
The IoBT system aims to incorporate a visualisation component accessible to authorised personnel, who can then effectively interpret the information and make informed decisions in real time.
Furthermore, the CAPSARII IoBT system will investigate the integration of the Head-Up Display (HUD) or other media channels as key elements to provide real-time feedback to soldiers by leveraging audiovisual and tactile stimuli to convey critical information while minimising cognitive load.
For example, the HUD displays interpretable visual indicators for vital signs, including heart rate, respiratory rate and stress levels, relevant environmental factors, such as temperature, humidity and concentration of pollutants, using intuitive icons or colour changes.
Simultaneously, reliable and robust haptic stimuli, such as vibrations, can provide immediate feedback corresponding to changes in processed information.
These stimuli, taking into account their frequency and amplitudes, can be convenient and be used in situations where audio or video feedback is not feasible.

The CAPSARII IoBT system is characterised by its full integration of wearable technologies, advanced data analytics, user-centric interfaces, and collaborative cloud-based strategies.
By combining these elements, CAPSARII aims to design a state-of-the-art solution that meets the demands of military operations and also fosters collaboration among European Defence.
The key components of the technical approach include: smart textiles and wearable system integration, the IoT architecture, the co-design of hardware and software, the AI-model optimisation, and the new generation of ultra-low power AI processors.

\subsection{Smart Textiles}

Smart textiles are fabric materials integrated with electronic components to provide advanced functionality, including sensing, actuation, communication and data processing. 
Recent advances in smart fabric technology include flexible and stretchable sensors and actuators, energy harvesting and storage, and miniaturised integrated electronics. 
The integration of intelligence into wearable textile systems that can process acquired data is creating a transformative framework, which drives the need to study more advanced devices, better connectivity, durability, safety and energy efficiency. 
Integration with IoBT systems, user training and ethical considerations are also key aspects of progress. There will also be a focus on improving ergonomics for military personnel with a multidimensional approach \cite{knight2002comfort}, promoting a circular economy by recycling and reusing materials in these devices, and effectively integrating wearable technologies into training and battlefield operations to improve safety and efficiency.

\subsubsection{Wearable System Integration}
Integration of advanced wearable devices with soldiers' gear, capturing real-time data on physiological, movement and environmental parameters, and operational metrics, by studying the integration and embedding of state-of-the-art sensors into wearable devices. Particular focus will be put into eye-tracking with an event-based camera that can perform similarly to high-end commercial detectors.
Furthermore, the wearable devices provide soldiers and commanding officers with intuitive interfaces for data submission, visualisation, and two-way communication, focusing on reducing cognitive overload.
Parametric design principles (specific for operational theatres and roles), allow for the customization of wearable devices, data analytics, and visualisation interfaces.
Integration with the soldier's gear needs to take into account that the devices will be compliant with necessary safety and reliability standards (e.g., the ATEX directive), such as the standards for use in potentially explosive environments, like aircraft refuelling stations and hangars.

\subsection{IoT Architecture}

IoT architecture is a framework that defines the structure and components of an IoT system, illustrating how various devices and services interact to collect, process, and exchange data through secure mechanisms, in compliance with regulations on data sovereignty. It typically includes several layers, each with specific functions \cite{zhong2017study}.
The \emph{perception layer} is where the sensors and actuators are, which in CAPSARII is embedded into the wearable devices.  The \emph{network layer} is where the data collected from the perception layer is transmitted and sent to the central processing unit,
with some basic preprocessing such as data aggregation, abstraction and analysis.  The \emph{application layer} is where data processing is carried out, providing insight, monitoring and basic automation.
IoT applications and user interfaces reside in the latter layer, enabling users to access and interact with the IoT system. This could be through web applications, mobile apps, or other interfaces. 
CAPSARII will design an IoT architecture for the IoBT system with a device layer consisting of wearables, based on a mesh network, with dynamic master nodes, an edge node with AI capabilities to make low latency decisions, and a server layer using different cloud platforms.
The IoT architecture should facilitate efficient flow of data from wearable devices to external servers for analysis and system design to efficiently scale based on demand and incorporate redundancy measures for system resilience.  The architecture should include edge computing for initial data processing, secure transmission protocols, and cloud-based federated analytics for in-depth evaluations and pattern mining, minimising latency and improving real-time decision making through initial edge data processing and analysis. 

\subsection{Body Area Network}

A BAN, also known as a Body Sensor Network (BSN), is a wireless network that connects small, low-power sensors and devices located on or near the human body. These networks utilise wearable sensors and short-range wireless communication technologies to transmit data to external devices, such as smartphones or fixed monitors. One applicable protocol for BANs is Thread \cite{unwala2018thread}, which is an open standard, wireless networking protocol that is designed for low-power, wireless IoT devices. It is built on top of the IEEE 802.15.4 standard and provides a reliable and secure way for devices to communicate in a mesh network. Thread is intended for home automation and other IoT applications where low power consumption, scalability, and reliability are essential. 
It features mesh networking with self healing networks that can recover when nodes get unavailable, IPv6 support, integrated security capabilities, low power which is crucial for battery-operated IoT devices and interoperability. The Zenoh protocol \cite{zenoh} is an option to lower the latency and increase the throughput when used on top of Thread.
We aim at evaluating new security mechanisms to improve authentication, encryption and integrity of messages exchanged within the BAN in order to mitigate risks associated with potential cybersecurity attacks on edge devices and networks too. In addition, we aim to reduce the data transmission latency, while increasing the network throughput, but maintaining or improving energy efficiency, frequency of security breaches, device authentication speed, data integrity rate and overall network uptime.

\subsection{Security}

\subsubsection{Device Identity and Authentication}

In today's rapidly evolving landscape of digital connectivity, managing the identities of end devices presents a multifaceted challenge. The scalability of identity management systems must keep pace with the explosive growth of connected devices, all while maintaining secure authentication processes. However, this must be accomplished without compromising user privacy. Furthermore, the security of these systems is paramount, as they themselves can become targets for malicious actors. Ensuring seamless interoperability between devices with different identity standards adds another layer of complexity. Managing the entire lifecycle of device identities, from provisioning to decommissioning, is essential yet often overlooked.
Achieving a balance between these challenges is critical for safeguarding the integrity and confidentiality of digital identities in our interconnected world.
The objective is to design an efficient identity management platform that takes into account the requirements of the end devices involved in this project. This will encompass considerations of the heterogeneity and resource limitations of these end devices, all while ensuring the high levels of security and privacy demanded in military environments.  Furthermore, we will also address the emerging threats posed by quantum computing~\cite{nist-pq}.


\subsubsection{Endpoint Security for RISC-V}

There is a rising interest in RISC-V security, as demonstrated by the activities of the RISC-V International Security Standing Committee, which include cryptographic extensions and trusted execution environments~\cite{lu2021survey}.
While these activities concern primarily established primitives (e.g., AES, SHA-2), recent competitions~\cite{caesar,nist-lc} have put forward strong candidates for lightweight, yet secure, encryption primitives that are necessary to ensure ultra-low-power operation.
CAPSARII focuses its endpoint security activities at the level of the microcontroller, where the interface between the operation of the platform and the external world resides. In particular, we aim at exploring the interaction between lightweight cryptography, hardware acceleration, and defence against side channels and covert channels~\cite{agosta2020compiler}.

\subsection{AI Models}

AI models will be developed and combined with non-AI models to extract relevant information from the acquired data. Specifically, raw data will be processed to extract parameters (e.g., pulse rate extracted from photoplethysmography sensor waveform), and obtained parameters will be considered to detect states (e.g., operator's fatigue) or provide synthetic indicators (e.g., air quality). Both machine and deep learning models will be chosen adequately to deal with specific data formats, for instance time series.

\subsubsection{Learning}

Transfer learning enables training a model on one task and then repurposing it for a related task, saving computational time and resources while often improving performances, especially with limited data. Two popular transfer learning approaches are fine-tuning, where the model is adjusted with new task data, and feature extraction, where the backbone of the first model serves as a feature extractor for the new task.
Meta learning extends this by enabling models to adapt to new tasks without forgetting previous ones, addressing the issue of catastrophic forgetting. In parallel, federated learning~\cite{chu2022design, lan2023elastically} offers a decentralized training method where multiple devices collaboratively build a shared model without transferring local data, making it suitable for privacy-sensitive and distributed scenarios~\cite{WANG2023103167,RANI2023110658}.
A federated meta learning approach will be designed in the CAPSARII project, validated on experimental data collected from human participants during trials designed specifically for the project~\cite{alday2020classification,singstad2020convolutional}, capable of predicting health and psychological conditions for each user, and of progressively adapting to different scenarios in the battlefield.

\subsubsection{Optimisation for Efficiency}

In recent years, AI has gained significant attention for its broad impact across sectors, with applications such as image classification and object detection~\cite{pouyanfar2018survey}. To improve accuracy, models — especially in deep learning — have become increasingly complex, often comprising millions or billions of parameters. This complexity poses challenges for deployment on resource-constrained devices like smartphones or IoT sensors~\cite{li2018deep}, which face limitations in memory, storage, and energy compared to data center servers~\cite{Chen2019Deep}. While cloud computing offers greater processing power, it introduces latency and privacy concerns due to the physical distance and data transmission risks.
In CAPSARII, we use a combination of model compression techniques to optimise AI models for deployment on resource-constrained wearable systems. A typical approach is pruning, which removes unnecessary neurons or connections from a trained model to reduce its size while preserving performance~\cite{han2015deep}. Quantization lowers the precision of weights and activations, reducing memory and computation needs~\cite{courbariaux2015binaryconnect}. Other techniques include knowledge distillation, where a smaller student model mimics a larger teacher~\cite{hinton2015distilling}, and weight sharing, which reuses parameters across the network~\cite{ullrich2017soft}. Together, these methods support the deployment of robust, low-power AI models on wearable devices in CAPSARII.

\subsection{Ultra-Low-Power Platforms for Edge AI}

When dealing with applications, such as Edge AI, that are characterised by extremely tight constraints in terms of energy, power, and size with respect to the computational requests, it is necessary to precisely tailor both hardware and software to the application requirements.
Thus, in embedded systems design, the methodologies of HW/SW co-design and platform-based design have gained recognition by allowing the exploration of a range of possible solutions based on configurable (and possibly reconfigurable) hardware platforms~\cite{demicheli2013hardware}.
CAPSARII proposes an HW/SW co-design methodology centred around a common platform element, the secure RISC-V microcontroller, which supports the execution of control code, communication, and duty cycling aspects of the application.
Then, a range of possible accelerators, both based on Processing-in-Memory (PIM) and GPU-like, can be selected to obtain the right configuration for each application.




\subsubsection{Processing-in-Memory}
PIM is an emerging compute paradigm to overcome memory-bound issues like bandwidth limitation, long data access latency, and very high data access energy, which are encountered in conventional hardware accelerators and compute platforms designed for memory-intensive applications.
Memory can be exploited for computation by either designing architectures where compute elements are located near the memory (e.g., exploiting 3D stacking) or by exploiting the physical properties of the chip materials to perform operations~\cite{asifuzzaman2023survey}.
%
Researchers have investigated various memory types as PIM candidates, particularly for the deployment of neural networks.
Such workloads are particularly appropriate because of their reliance on matrix and matrix-vector multiplications, which can be efficiently implemented with PIM techniques, and because the network weights are generally constant, allowing an implementation in RRAM and non-volatile memories~\cite{ielmini2018memory}.
Major memory vendors like Samsung, and start-ups like UPMEM are researching on PIM and the first engineering samples of DRAM-based PIM accelerators were published in \cite{kwon202125} and \cite{devaux2019true}, respectively.
CAPSARII will explore the use of such technologies to speed up and improve the energy efficiency of the execution of its ML tasks.

\subsubsection{Precision Tuning}
Error-tolerating applications are increasingly common, particularly among AI-based ones.
Proposals have been made at the hardware level to take advantage of inherent perceptual limitations, redundant data, or reduced precision input, as well as to reduce system costs or improve power efficiency.
At the same time, works on floating-point to fixed-point conversion tools allow us to trade-off the algorithm exactness for a more efficient implementation~\cite{cherubin2020survey}.
Finally, new data types (e.g., bfloat16, posits) are emerging, and post-Von Neumann computing models promise massive improvements in energy efficiency.
We aim at extending the tools for precision tuning developed as part of the H2020 FETHPC ANTAREX project~\cite{8344645} to cover emerging architectural solutions, including PIMs and embedded GPUs.
These tools, collected in the TAFFO (Tuning Assistant for Floating-Point to Fixed-Point Optimization) framework~\cite{Cattaneo2022TAFFO} are implemented as a set of plugins for the LLVM compiler, and, based on programmer hints expressed, perform value range analysis, data type and code conversion, and static estimation of the performance impact.
We aim at improving the performance estimation through a deeper characterization of the execution element, and by considering emerging data types and innovative architectural solutions for AI acceleration~\cite{le2018mixed}.



\section{Conclusion}

The CAPSARII project aims at developing an innovative IoBT wearable system and related ICT framework for supporting military staff.  Tactical decision support will be enabled through monitoring physiological, movement and environmental parameters and by employing AI models deployed on the edge nodes.  Two use cases will be studied: 
\begin{inparaenum}
\item Pilot fatigue, where the wearable system could for instance provide awareness of a crew member's fatigue level and suggest personalised countermeasures;
\item Warfighters in harsh environments, where the health status of individual soldiers and the whole team can be monitored and analysed, e.g. to prevent stress and fatigue-related accidents. 
\end{inparaenum}
 Through these two use cases, CAPSARII will provide research into wearable systems, HW/SW co-design, efficient AI models at the edge, security aspects of IoBT and low power devices. 

\bibliographystyle{IEEEtran}
\bibliography{refs}

\end{document}